\documentclass[11pt,twocolumn,A4]{article}
\usepackage[dvips]{graphicx}
\usepackage{setspace}
\begin{document}
\onecolumn

\begin{center}
{\bf{\Large Controlled electron transport in coupled Aharonov-Bohm rings}}\\
~\\ 
Supriya Jana and Arunava Chakrabarti$^\dag$ \\
~\\
{\em Department of Physics, University of Kalyani, Kalyani,  
West Bengal-741 235, India.} \\
~\\ 
{\bf Abstract}
\end{center}
We propose a simple model of two coupled mesoscopic rings threaded by 
magnetic flux which mimics a device for electron transmission in a 
controlled fashion. Within a tight binding formalism we work out exactly the 
conditions when a completely ballistic transmission can be triggered at any desired 
position of the Fermi level within a continuous range
of energy. The {\it switching} action can be easily controlled  
by appropriately 
tuning the values of the flux through the two rings. The general transmission 
across the ring system displays interesting features characterized by the 
appearance and collapse of Fano resonances, and the supression of the Aharonov-Bohm 
oscillations in several casess. We report and analyze exact analytical results for a 
few such cases. 
\vskip 0.4cm
\begin{flushleft}
\end{flushleft}
\vskip 0.3cm
\begin{flushleft}
{\bf $^\dag$Electronic mail}: arunava\_chakrabarti@yahoo.co.in
\end{flushleft}
\newpage
\noindent
{\bf 1. Introduction}
\vskip .2in
The study of electronic conductance in mesoscopic rings threaded by a magnetic 
flux has been an important and extensively studied field for many years
~\cite{byers}-\cite{amato}. With the experimental realization of mesoscopic 
rings and cylinders~\cite{sharvin}-\cite{gijs} and later, the quantum dots 
and quantum wires~\cite{mitin}-\cite{ferry}, the research in this field 
got the desired momentum. Present day nanofabrication and lithographic 
techniques have enabled us to 
study the effects of quantum coherence in small scale systems and to explore 
the possibility of controlling electronic transport in tailor made devices
~\cite{sundar}-\cite{gambar}.   

In recent years one witnesses a large number of theoretical works which focus on the 
basics of quantum transport in low dimensional systems using very simple tight binding 
models~\cite{vasseur}-\cite{rod}. The models find their justification in the 
state of the art mesoscopic and 
nanotechnology where using the tip of a scanning tunnel microscope (STM) one can 
literally build up an atomic wire~\cite{pou,rod1} or an array of quantum dots with a 
desired geometry. 
Interestingly enough, even studies with non-interacting electrons have been found to 
produce results which in many cases bring out the essential qualitative features 
of the real-life systems and at the same time 
throws light into the basics of quantum transport in quasi-one dimension~\cite{shi}-
\cite{torres}.

In this communication we address the problem of electron transport across a 
model mesoscopic ring (the Primary ring) which is coupled at one point 
to a `secondary' ring. Both the rings contain magnetic flux perpendicular to 
the plane of the rings. 
The basic motivation behind the present work is to look for a comprehensive control 
over the transmission across a mesoscopic ring by an atomic cluster attached to 
it from one side. To our mind, inspite of the existing studies there is ample scope 
to explore the role of an external magnetic field in gaining control over  
electronic transmission in 
closed loop geometries, and in designing a switching device 
even using rings with minimal sizes. We get quite inspiring results. For example,
we show that even with a minimum number of atomic sites 
contained in the rings, the system offers a very efficient way of controlling 
electron transport that can be tuned at will by mutually adjusting the magnetic 
field and the inter-ring coupling. In particular, a sharp {\it switch action} can be triggered  
at {\it any position of the Fermi level} within a given range. 
This can be quite important from the standpoint 
of designing a quantum tunnel device. In addition to this, interesting Aharonov-Bohm (AB) 
oscillations~\cite{aharon} as well as the supression 
of AB oscillations as a function of the inter-ring coupling parameter are observed. Fano lineshapes 
~\cite{fano} are observed in the transmission spectrum which can be made to collapse 
and even disappear by controlling the secondary flux. 

In what follows we present the results of our analysis. In Section 1 we describe the 
model. The switch action and the transmission profiles are presented in section 2. 
In section 3 we analyse the appearance of the Fano lineshapes and in Section 4 we 
draw conclusions.
\vskip .25in 
\noindent
{\bf 2. The Model and the method}
\vskip .25in

Let us begin by refering to Fig.1(a). The `primary' ring (P) contains a total 
of $l+2m+1$ sites and is coupled to the secondary ring (S) which contains $N+1$ 
sites. The contact point is marked by the site $\mu$, and the P-S hopping integral 
is $\lambda$. A magnetic flux $\Phi_P$ threads the primary ring, while the flux 
through the secondary ring is $\Phi_S$. Two semi-infinite perfect leads are 
connected to the primary ring at the end-points $A$ and $B$ as shown. The hamiltonian, 
in the tight binding approximation, for the lead(L)-ring-lead system is given by,
\begin{equation}
H = H_{L} + H_{P} + H_{S} + H_{P-S} + H_{P-L}
\end{equation}
where,
\begin{eqnarray}
H_{L} & = & \epsilon_0\sum_{i=-\infty}^{A-1} c_i^{\dag}c_i + 
\epsilon_0\sum_{i=B+1}^{\infty} c_i^{\dag}c_i + 
t_0 \sum_{<ij>} c_i^{\dag} c_j + h.c. \nonumber \\
H_{P} & = & \epsilon_A c_A^{\dag}c_A + \epsilon_B c_B^{\dag}c_B +
\epsilon_0 \left (\sum_{i\in \cal U} c_{i,P}^\dag c_{i,P} + 
\sum_{i \in \cal L} c_{i,P}^\dag c_{i,P} \right ) +  
t_0 \exp(i\theta_P)\sum_{<ij>} c_{i,P}^{\dag}c_{j,P} + h.c. \nonumber \\
H_{S} & = & \epsilon_0 \sum_{i=1}^{N} c_{i,S}^{\dag}c_{i,S} +  
t_0 \exp(i\theta_S)\sum_{<ij>} c_{i,S}^{\dag}c_{j,S} + h.c. \nonumber \\
H_{P-S} & = & \lambda c_{\mu,\cal L}^{\dag}c_{1,S}  \nonumber \\
H_{P-L} & = & t_0 (c_A^{\dag}c_{A-1} + c_B^{\dag}c_{B+1}) 
\end{eqnarray}
In the above, $c_{i}^{\dag}(c_i)$ represents the
creation (annihilation) operator for the leads, the ring and the QD chain at the appropriate sites.
The symbols $P$ or $S$ in Eq.(1) as well as in the subscripts in Eqs.(2)  
represent the primary or the secondary ring in repective order. 
$\cal U$ and $\cal L$ symbolize the upper and the lower arms of the primary rings.  
order.
The on-site potential at the leads, at the edges $A$ and $B$, and in the bulk of the rings
are all taken to be $\epsilon_0$ including the site marked $\mu$. 
So, we are primarily interested in the effect of the geometry on the transport properties.
The amplitude of the hopping integral
is taken to be $t_0$ throughout except the hopping from the site $\mu$ in the ring to the first site
of the QD chain, which has been symbolized as $\lambda$ and represents the `strength' of coupling
between the rings P and S. $\theta_{P}$ and $\theta_S$ are given by
$\theta_{P} = 2\pi \Phi_{P}/(l+2m+1)\Phi_0$, and $\theta_s = 2\pi \Phi_{S}/(N+1)\Phi_0$
respectively, where, 
$\Phi_{P(S)}$ is the flux threading the primary (secondary) ring and $\Phi_0=hc/e$ is
the fundamental flux quantum.
The task of solving the Schr\"{o}dinger equation to obtain the
stationary states of the system can be reduced to an equivalent problem of solving a set of
difference equations. We list the relevant equations below.

For the sites $A$ and $B$ at the ring-lead junctions,
\begin{eqnarray}
(E-\epsilon_A) \psi_A & = & t_0e^{i\theta_P} \psi_{1,\cal U} + 
t_0e^{-i\theta_P} \psi_{1,\cal L} + t_0 \psi_{A-1} \nonumber \\
(E-\epsilon_B) \psi_B & = & t_0e^{-i\theta_P} \psi_{l,\cal U} + t_0e^{i\theta_P} \psi_{2m+1,\cal L} + 
t_0 \psi_{B+1} 
\nonumber \\
\end{eqnarray}
In the above, $\psi_{A-1}$ and $\psi_{B+1}$ represent the amplitudes of the wave function at the sites on the
lead which are closest to the points $A$ and $B$, and,
$\cal U$ and $\cal L$ in the subscripts refer to the `upper' and the `lower' arms of the 
primary ring respectively.

For the sites in the bulk of the primary ring the equations are,
\begin{eqnarray}
(E-\epsilon_0) \psi_{j,\cal U} & = & t_0e^{-i\theta_P} \psi_{j-1,\cal U} + t_0e^{i\theta_P} \psi_{j+1,\cal U} \nonumber \\
(E-\epsilon_0) \psi_{j,\cal L} & = & t_0e^{i\theta} \psi_{j-1,\cal L} + t_0e^{-i\theta_P} \psi_{j+1,\cal L} 
\end{eqnarray}
and, 
for the site marked $\mu$ in the lower arm of the ring, the equation is,
\begin{equation}
(E-\epsilon_0) \psi_{\mu, \cal L} = t_0 e^{i\theta_P}\psi_{\mu-1,\cal L} + t_0 e^{-i\theta_P}\psi_{\mu+1,\cal L}
+ \lambda \psi_{1,S}
\end{equation}
$\mu \pm 1$ implying the sites to the right and to the left of the site marked $\mu$ respectively. Finally,
for the sites in the secondary ring  we have the following set of equations,
\begin{eqnarray}
(E-\epsilon_0) \psi_{1,S} & = & \lambda \psi_\mu + t_0 e^{i\theta_S} \psi_{2,S} 
+ t_0 e^{-i\theta_S} \psi_{N,S} \nonumber \\
(E-\epsilon_0) \psi_{2,S} & = & t_0 e^{i\theta_S} \psi_{3,S} + t_0 e^{-i\theta_S} \psi_{1,S} \nonumber \\
. \nonumber \\
. \nonumber \\
(E-\epsilon_0) \psi_{N,S} & = & t_0 e^{i\theta_S} \psi_{1,S} + t_0 e^{-i\theta_S} \psi_{N-1,S}
\end{eqnarray}
where the central set of equations above refer to the bulk sites viz. $j=2,...N-1$ in the QD array

To calculate the transmission coefficient across such a coupled ring system we 
adopt the following steps. First, the dangling S-ring is `wrapped' up into an effective 
site by decimating the amplitudes $\psi_{2,S}$ to $\psi_{N,S}$ in the secondary ring.
The renormalized on-site potential of the first site of the S-ring is given by,
\begin{equation}
\epsilon^* = \epsilon_0 + \frac{2t_0^2 (E-\epsilon')}{\Delta} + \frac{2t_0^3}{\Delta U_{N-2}(x)} 
\cos (2\pi\Phi_S/\Phi_0)
\end{equation}
where,
\begin{eqnarray}
\epsilon' & = & \epsilon_0 + 
\frac{t_0U_{N-3}(x)}{U_{N-2}(x)} \nonumber \\
\Delta & = & (E-\epsilon')^2 - \frac{t_0^2}{U_{N-2}^2(x)}
\end{eqnarray}
In the above equations, $x=(E-\epsilon_0)/2t_0$ and $U_j(x)$ is the $j$th order Chebyshev
polynomial of the second kind, with $U_0=1$ and $U_{-1}=0$.
The bulk sites of the primary ring are then decimated to convert the P-S ring system 
into an effective `diatomic molecule' $A-B$ clamped between the leads (Fig.1(b)). The 
effective on-site potentials at $A$ and $B$ and the renormalized flux dependent 
hopping integral between them are given by~\cite{jana},
\begin{eqnarray}
\sigma_{A(B)} & = & \epsilon_0 + t_0 \left [ \frac{U_{l-1}}{U_l} + \frac{U_{m-1}}{U_m} \right ] + 
\left ( \frac{t_0}{U_{m}} \right )^2 \chi (E,\lambda,\Phi_S) \nonumber \\
t_F & = & \frac{t_0}{U_l} e^{i\theta_P(l+1)} + \frac{t_0^2}{U_m^2} \chi (E,\lambda,m)
e^{-i(2m+2)\theta_P}
\end{eqnarray}
where, $\chi(E,\lambda,m) = [E - \omega -2t_0 U_{m-1}/U_m]^{-1}$, with 
$\omega = \epsilon_0 + \lambda^2/(E-\epsilon^*)$. The two-terminal transmission coefficient 
across the primary ring is given by~\cite{stone}, 
\begin{equation}
T = \frac{4 \sin^2 qa}{|M_{12} - M_{21} + (M_{11} - M_{22}) \cos qa|^2 + 
|M_{11} + M_{22}|^2 \sin^2 qa}
\end{equation}
Here, $a$ is the lattice spacing in the leads, taken as unity throughout. 
$q = \cos^{-1}[(E-\epsilon_0)/2t_0]$ is the wave vector. The Transfer matrix elements  
are given by, 
$M_{11} = (E-\sigma_A)(E-\sigma_B)/(t_0t_F) - t_B/t_0$; $M_{12} = -(E-\sigma_B)/t_F$; 
$M_{21} = (E-\sigma_A)/t_F$, and $M_{22} = -t_0/t_F$.
\vskip .3in
\noindent
{\bf 3. Results and discussions}
\vskip .3in
\noindent
{\it (a) The switch action} 
\vskip .25in 
We first discuss the principal result of the present work, i.e. how can such a system 
of coupled rings be used to trigger ballistic transmission at any desired energy within a 
given range. To make things less obscure, we concentrate on the simplest situation where, 
the primary (P) ring contains four sites (the sites at $A$, $B$, and one site in each arm), 
and the secondary (S) ring has just three sites. We call such a system a $(4,3)$ system. 
The effective on-site potential $\epsilon^*$ in this case is given by, 
\begin{equation} 
\epsilon^* = \epsilon_0 + \frac{2t_0^2 (E-\epsilon_0)}{\Delta} + \frac{2t_0^3}{\Delta} \cos 3\theta_S
\end{equation}
where, $\Delta = ( E - \epsilon_0 )^2 - t_0^2$.
It is known~\cite{jana} that, for a primary flux $\Phi_P = \Phi_0/2$, such a system 
exhibits a ballistic transmission at $E = \epsilon^*$ which, in the present case leads to an 
equation, $\cos 3\theta_S = F(E)$, where, 
\begin{equation}
F(E) = \frac{E-\epsilon_0}{2t_0^3} \left [(E-\epsilon_0)^2 - t_0^2 \right ]
\end{equation}

Eq.(12) implies that, for the entire range of energy for which $F(E)$ remains 
bounded by $\pm 1$, one can easily fix up a value of the secondary flux $\Phi_S$ such that 
the transmission across the primary ring is unity, as discussed elsewhere~\cite{jana}. 
In the simple case of the $(4,3)$ system, the function $F(E)$ remains within $\pm 1$ 
for all values of the energy $E$ within $(\epsilon_0 - 2t_0, \epsilon_0 + 2t_0)$, 
i.e. for the entire allowed band of the semi-infinite ordered leads. This is illustrated 
in Fig.2. Thus, the $(4,3)$ system can be tuned by the secondary flux to act as a 
switch corresponding to any desired position of the Fermi level within the said range.
For example, with $\epsilon_0 = 0$ and, $t_0 = 1$, if we set the Fermi level at $E = 1$, 
then a ballistic transmission $(T = 1)$ is obtained by setting $\Phi_S = \Phi_0/2$. 
The transmission at all other values of the secondary flux remains zero. This is what we 
mean by a {\it switch action}. For energy values other than $E = 1$ we get a couple of 
values of $\Phi_S$ placed symmetrically around $\Phi_S = \Phi_0/2$. Interestingly, 
as one approaches the value $E = 1$, the separation between the two values of the 
secondary flux which correspond to the resonant transmission get reduced smoothly, and 
finally, at $E = 1$ the pair of peaks collapse to give a single sharp resonance peak in 
the transmission spectrum. This aspect is depicted in Fig.3. The appearance of a single 
resonance peak at $E = 1$ can be understood easily by analyzing the expression of $\epsilon^*$.

Before we end this section it should be pointed out that, the switch action is strongly 
dependent on making the effective hopping across the vertices $A-B$ in the primary ring.
This can be achieved whenever one has an equal number of scatterers in the upper and the 
lower arms of the primary ring, and sets $\Phi_P = \Phi_0/2$. We have also investigated 
many cases when the secondary ring contains more than just three atoms. As long as the selected 
energy (the Fermi energy) is chosen to be within the spectrum of the secondary ring (and 
also within the allowed band of the leads), there will be a switch action. Hence, the 
result presented for the $(4,3)$ case is perfectly general, and really can be 
achieved independent of the size of the rings when one tunes the hamiltonian parameters 
appropriately.
\vskip .25in
\noindent
{\it (b) Fano lineshape in the transmission spectrum:}
\vskip .25in
Fano lineshape~\cite{fano} in the transmission is widely displayed by several 
mesoscopic systems~\cite{rod}. This typical resonance profile is observed when 
a discrete, bound level get `mixed' with a continuum of states. Atomic clusters dangling 
from a linear chain provide such an example~\cite{mirosh,arun}, and the present case 
is no exception. However, in this case interesting changes in the Fano lineshapes in 
the transmission spectrum take place as the flux through the secondary ring is varied 
keeping the primary flux fixed. The opposite is also true, but we report just 
one case, viz, the effect of variation of the seconadry flux, to save space. To this 
end, we have analyzed in some details the transmission amplitude for the $(4,3)$ system 
for particular values of the energy $E$ in the $\delta$-neighborhood of a transmission 
minimum (as the secondary flux is varied). The transmission amplitude, apart from a 
phase factor, is given by, 
\begin{equation}
\tau \sim 2 \sin q \frac{G_2 \delta + \frac{F_2}{G_2}}{F_4 (\delta + \frac{G_4}{F_4}) + 
i F_5 (\delta + \frac{G_5}{F_5})}
\end{equation}
where, $F_5 = -F_2F_3 + G_2G_3 + 2G_1 \cos q$; $G_5 = G_3F_2 - 2 (E-F_1) \cos q - t_0$; 
$F_4 = (G_3G_2 - F_3F_2 - G_1) \sin q$; $G_4 = (E - F_1 + F_2G_3) \sin q + t_0 \sin 2q$;
$F_3 = [2G_1(E-F_1) + G_2^*]/t_0$; $G_3 = [(E-F_1)^2 - F_2^*]/t_0$; 
$F_2 = t_0^2 \exp(2i\theta_P)/(E-\epsilon_0) + x \exp(-2i\theta_P)/y$; 
$G_2 = (z - b/(E-a)) x \exp (-2i\theta_P)/y$; 
$F_1 = \epsilon_0 + t_0^2/(E-\epsilon_0) + x/y$; $G_1 = x (z - b/(E-a))/y$; 
$x = t_0^2 (E-a)$; $y = (E - \epsilon_0)(E - a) - \lambda^2$ and $z = b (E - \epsilon_0)/y$.

The asymmetry parameter~\cite{arun,voo} is given by $F_2/G_2$, which is, in general, 
complex in the present geometry. The numerator of Eq.(13) is zero when $\delta = -F_2/G_2$, 
while the real part of the denominator vanishes for $\delta = -G_4/F_4$. These 
{\it de-tuned} zeros give rise to an asymmteric Fano profile~\cite{voo} around the 
secondary flux $\Phi_S = \Phi_S^{min}$, $\Phi_S^{min}$ is the secondary flux 
corresponding to the transmission minimum in Fig.4. The sign of the quantity 
$Q = (F_2/G_2 - G_4/F_4)$ is responsible for whether the Fano lineshape exhibits a 
{\it peak to dip} drop in the transmission or a sharp {\it dip to peak} rise in it.
It is simple to check that the sign of $Q$ changes as one the secondary flux 
from $\Phi_S = \Phi_0/2 - \Phi_S^{min}$ to $\Phi_S = \Phi_0/2 + \Phi_S^{min}$. 
This swapping of the Fano profile is clearly visible in Fig.4 where we have plotted 
the transmission coefficient as a function of the secondary flux when the primary 
flux is quite arbitrarily fixed $\Phi_P = \Phi_0/5$. We have taken the P-S coupling 
$\lambda = 0.08 t_0$, a weak one. For strong coupling, the Fano resonances become broadened 
as $\lambda$ controls the width of resonance. As the energy is gradually changed from 
$E = 0$ to $E = 1$, interestingly we observe that, while for $E = 0$ the Fano like 
behaviour is obscure (the transmission coefficient appears as delta-like peaks only), 
with increasing energy the asymmetric Fano profile is clearly visible and the lineshape 
swaps from a {\it dip-to-peak} to a {\it peak-to-dip} pattern as the secondary flux 
crosses the half flux quantum mark. In addition to this, the gap between the Fano anti-resonances 
shrinks as one approaches the value $E = 1$, and finally disappears completely at $E = 1$ (not 
shown in this picture). This latter observation is easily justified from Eq.(11). The features 
exhibited in Fig.4 are also observed even if we set the primary flux equal to zero.

The effect of the P-S coupling $\lambda$ on the transmission spectrum has also been 
studied in details. Apart from controlling the width of the lineshape, an increasing 
$\lambda$ can suppress the AB oscillations as shown in Fig.5. Here, we present the 
results for $E = 0$ and the primary flux set equal to the half flux quantum. The overall 
transmission enhances as $\lambda$ is increased, reaching alsmost unity as the 
inter-ring coupling becomes sufficiently large. This gives us an idea of the effect of 
the proximity of the primary and the secondary rings.
\vskip .25in 
\noindent
{\bf 4. Conclusions}
\vskip .25in 
A coupled ring system can exhibit a wealth of transmission characteristics as the 
magnetic flux threading the rings are adjusted. We have focussed on certain specific issues, 
one of which is the possibility of a ballistic transport at any desired value of the 
Fermi energy that can be controlled by tuning the secondary flux. This gives the coupled 
ring system the status of a {\it mesoscopic switch} which can be put `on' at will.
Fano lineshapes are observed in the transmission spectrum under various occasions and 
we have analyzed one such case where the Fano profile swaps as the secondary flux 
changes. Finally, the collapse of the Aharonov-Bohm oscillations as the inter-ring 
coupling increases is also discussed. 

\newpage

\noindent
{\bf Figure Captions:}
\vskip 0.4in
\noindent
Figure 1: (a) Schematic representation of the coupled ring system, and 
(b) renormalization of the ring system into a `diatomic molecule'.

\vskip 0.4in
\noindent
Figure 2: $F(E)$ versus $E$ curve for the $(4,3)$ ring system. We have 
chosen $\epsilon_0 = 0$, $t_0 = 1$ everywhere, and $E$ is measured in unit of $t_0$.

Figure 3: Transmission across a $(4,3)$ coupled ring system with $\lambda = 0.2t_0$. 
Here, $E = 0.5$ (solid), $0.75$ (dashed) and $1.0$ (dotted) measured in unit of $t_0$.
$\Phi_P = \Phi_0/2$ and the other parameters are same as in the previous figures.

\vskip 0.4in
\noindent
Figure 4: Fano profiles in the transmission spectrum of a $(4,3)$ coupled ring system. 
We have taken $\Phi_P = \Phi_0/5$, $\lambda = 0.08 t_0$ and $E = 0$ (solid line), 
$0.2$ (dashed line)$ and $0.75$ (dotted line)$ mesured in unit of $t_0$. $\epsilon_0$ and 
$t_0$ are as in Fig.2.

\vskip 0.4in
\noindent
Figure 5: Collapse of the AB oscillations as the inter-rine coupling $\lambda$ is 
increased. We have fixed $\Phi_P = \Phi_0/2$ and $E = 0$. $\lambda = 0.25$ (solid), 
$1.0$ (dashed), $3.0$ (dotted) and $5.0$ (thick) in unit of $t_0$. $\epsilon_0$ and 
$t_0$ are chosen as zero and unity as before.

\newpage

\begin{figure}[ht]
{\centering\resizebox*{8cm}{6cm}{\includegraphics{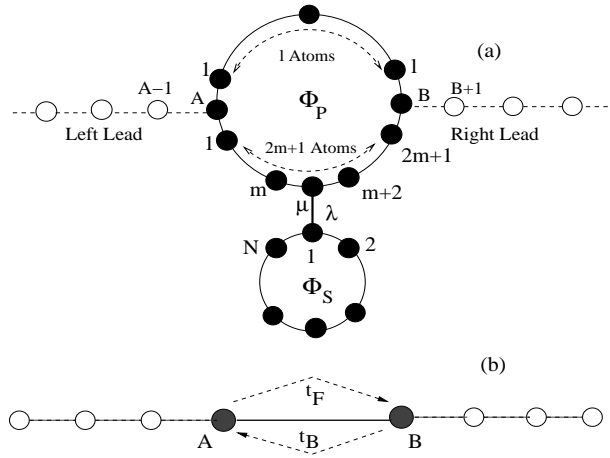}} \par}
\caption{ (a) Schematic representation of the coupled ring system, and 
(b) renormalization of the ring system into a `diatomic molecule'.} 
\end{figure}

\newpage
\begin{figure}[ht]
{\centering\resizebox*{8cm}{6cm}{\includegraphics [angle=0] {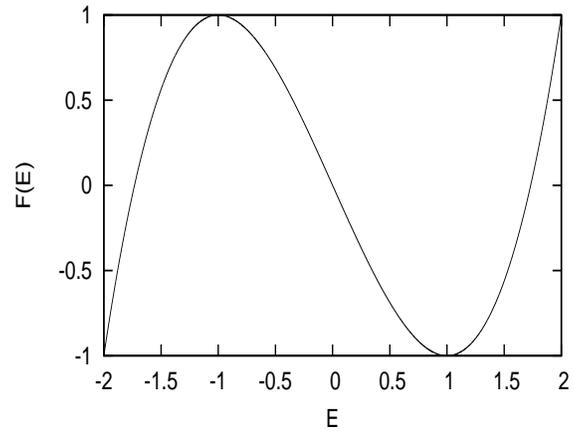}} \par}
\caption{
$F(E)$ versus $E$ curve for the $(4,3)$ ring system. We have 
chosen $\epsilon_0 = 0$, $t_0 = 1$ everywhere, and $E$ is measured in unit of $t_0$.}
\end{figure}

\newpage

\begin{figure}[ht]
{\centering\resizebox*{8cm}{6cm}{\includegraphics{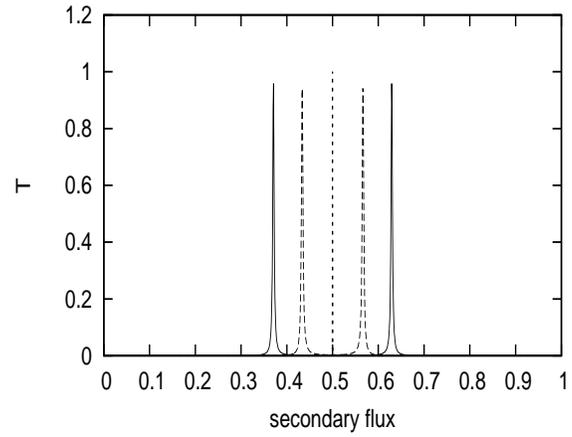}} \par}
\caption{
Transmission across a $(4,3)$ coupled ring system with $\lambda = 0.2t_0$. 
Here, $E = 0.5$ (solid), $0.75$ (dashed) and $1.0$ (dotted) measured in unit of $t_0$.
$\Phi_P = \Phi_0/2$ and the other parameters are same as in the previous figures.}

\end{figure}

\newpage

\begin{figure}[ht]
\begin{center}
\includegraphics[height=8cm,width=6cm,angle=-90]{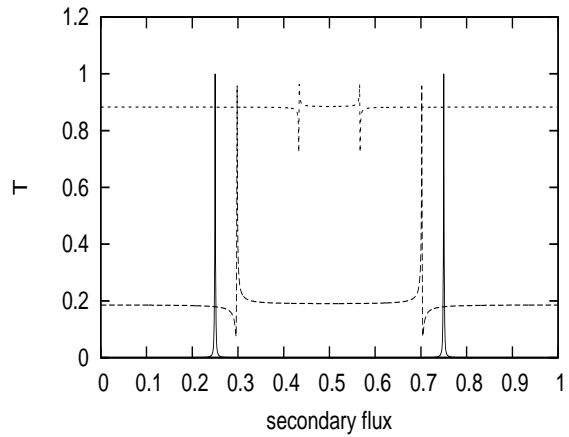}
\caption{ 
Fano profiles in the transmission spectrum of a $(4,3)$ coupled ring system. 
We have taken $\Phi_P = \Phi_0/5$, $\lambda = 0.08 t_0$ and $E = 0$ (solid line), 
$0.2$ (dashed line) and $0.75$ (dotted line) mesured in unit of $t_0$. $\epsilon_0$ and 
$t_0$ are as in Fig.2.}

\end{center}
\end{figure}

\newpage

\begin{figure}[ht]
\begin{center}
\includegraphics[height=6cm,width=8cm,angle=0]{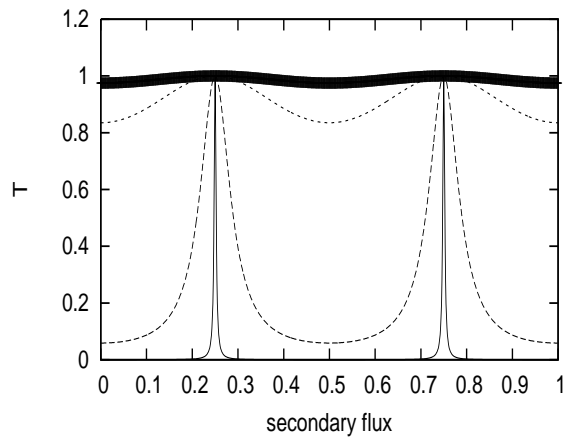}
\caption{
Figure 5: Collapse of the AB oscillations as the inter-rine coupling $\lambda$ is 
increased. We have fixed $\Phi_P = \Phi_0/2$ and $E = 0$. $\lambda = 0.25$ (solid), 
$1.0$ (dashed), $3.0$ (dotted) and $5.0$ (thick) in unit of $t_0$. $\epsilon_0$ and 
$t_0$ are chosen as zero and unity as before.}
\end{center}
\end{figure}

\end{document}